\RequirePackage{fix-cm}
\documentclass[smallextended]{svjour3}       % onecolumn (second format)
\smartqed  % flush right qed marks, e.g. at end of proof
\usepackage{graphicx,latexsym,amssymb,amsmath,color,mathrsfs,booktabs,ifpdf}
\usepackage{bm}% bold math
 \usepackage{mathptmx}      % use Times fonts if available on your TeX system
\usepackage{xspace}

\def\AA{nucleus-nucleus\xspace }
\newcommand{\OC}{\ensuremath{^{16}\mathrm{O}+^{12}\mathrm{C}}\xspace}
\newcommand{\OO}{\ensuremath{^{16}\mathrm{O}+^{16}\mathrm{O}}\xspace}
\newcommand{\CC}{\ensuremath{^{12}\mathrm{C}+^{12}\mathrm{C}}\xspace}
\newcommand{\CCC}{\ensuremath{^{13}\mathrm{C}+^{12}\mathrm{C}}\xspace}

\begin{document}
%\raggedbottom                     % không kéo giãn trang
%\setlength{\parskip}{3pt plus 2pt minus 1pt}   % khoảng cách đoạn vừa phải
%\setlength{\parindent}{2em}       % hoặc để 1em nếu bạn muốn thụt đầu dòng
\title{Nuclear Rainbow of Core-Symmetric Systems}

\author{Nguyen Tri Toan Phuc \and Nguyen Hoang Phuc \and  Dao T. Khoa}

%\authorrunning{N. T. T. Phuc et al.} % if too long for running head

\institute{Nguyen Tri Toan Phuc \at
	Department of Nuclear Physics, Faculty of Physics and Engineering Physics, University of Science, 
	Ho Chi Minh City, Vietnam \\ Vietnam National University, Ho Chi Minh City, Vietnam \\ 
	\email{nttphuc@hcmus.edu.vn}
	\and 	Nguyen Hoang Phuc \at
 Department of Applied Physics, Faculty of Applied Science, Ho Chi Minh City University of Technology 
 (HCMUT), 268 Ly Thuong Kiet Street, Dien Hong Ward, Ho Chi Minh City, Vietnam \\
	Vietnam National University Ho Chi Minh City, Linh Xuan Ward, Ho Chi Minh City, Vietnam \\ 
	\email{nguyenhoangphuc@hcmut.edu.vn}
	\and Dao T. Khoa \at
	Institute for Nuclear Science and Technology, VINATOM, 
	179 Hoang Quoc Viet Road, Nghia Do, Hanoi, Vietnam \\
	\email{daotienkhoa@icloud.com}}

\date{Received: date / Accepted: date}

\maketitle

\begin{abstract}
The nearside-farside (NF) decomposition method developed originally by Fuller for elastic scattering 
of a nonidentical \AA system was generalized to study the nuclear rainbow pattern in a symmetric 
or core-symmetric dinuclear system. It has been shown that the projectile-target identity of an identical 
system implies a symmetric interchange of the nearside and farside components of elastic scattering 
amplitude around $\theta_{\rm c.m.}=90^\circ$. A similar interchange appears also in a nonidentical 
core-symmetric system due to elastic transfer of cluster or nucleon between two identical cores. 
The analysis of the \CC, \OC, and \CCC systems shows how the generalized NF decomposition 
method reveals the nuclear rainbow pattern in these systems, which can be helpful in probing
the real optical potential and nuclear clustering.

\end{abstract}

\section{Introduction}
\label{sec:intro}
The elastic scattering of strongly bound light heavy ions exhibits rich diffractive and refractive structures, 
including the nuclear rainbow, associated with a broad Airy oscillation pattern of scattering cross 
section at medium and large angles \cite{Bra96,Bra97,Kho07r}. The observed nuclear rainbow 
was proven to be sensitive to the real \AA optical potential (OP) at small internuclear distances 
and, consequently, be used to probe density dependence of an effective nucleon-nucleon 
interaction through the folding model analysis \cite{Kho07r}. A useful tool for describing nuclear 
rainbow is the nearside-farside (NF) decomposition method originally developed by Fuller \cite{Ful75}, 
which splits the elastic scattering amplitude of a nonidentical \AA system ($A+B\to A+B$) into the 
waves deflecting from the near and far sides of scattering center. The interference between the refractive 
farside subamplitudes generated by a deep weakly-absorptive OP gives rise to a broad Airy oscillation 
pattern characteristic of nuclear rainbow in elastic $\alpha$-nucleus and light heavy-ion (HI) scattering 
\cite{Bra97,Kho07r}.

However, some interesting HI cases fall outside the scope of the Fuller decomposition 
method. In particular, the nuclear rainbow pattern could not be properly identified in elastic 
scattering data measured for identical or core-identical \AA systems, such as \CC and \OO or \OC 
and \CCC. Here, the projectile-target exchange or elastic transfer effects show up at large angles, 
and the rainbow pattern is distorted by the oscillation of scattering cross section caused by Mott 
interference around $\theta_{\rm c.m.}\simeq 90^\circ$ \cite{Phu18,Phu19,Phu21a,Phu24a}. 
In these cases, the boson symmetry requires the elastic scattering amplitude to be symmetrized,
which can be effectively taken into account by a parity-dependent OP \cite{Phu19}, and the 
original NF decomposition method by Fuller is no longer applicable for the separation 
of the nearside and farside scattering. 

We have, therefore, generalized Fuller's method to overcome this limitation, and extended the NF 
decomposition method to elastic scattering of light HI systems with core-exchange symmetry. 
An exact NF decomposition of the \emph{symmetrized} elastic scattering amplitude for two 
identical (spin-zero) nuclei into the direct and exchange NF subamplitudes revealed a characteristic 
``butterfly-wing'' exchange pattern of Airy minima around $\theta_{\rm c.m.}=90^\circ$ \cite{Phu24a}. 
The same formalism, together with a complex core-exchange potential given by the coupled reaction 
channel (CRC) calculation \cite{Phu19,Phu21a,Phu24a}, was applied to the core-identical systems 
such as \OC{} and, in preliminary form, \CCC{}. In these systems, elastic transfer of an $\alpha$ cluster 
or a nucleon must be included into the CRC calculation to properly account for both the anomalous 
large-angle scattering and deterioration of the Airy oscillation pattern. The core-exchange symmetry 
effect was also found in the inelastic $\alpha$ transfer that contributes to the inelastic \OC scattering 
at low energies \cite{Phu21c}. 

This paper presents the generalized NF decomposition method and illustrates 
its use in the description of the nuclear rainbow pattern observed in the identical and 
core-identical dinuclear systems. 

\section{Fuller Decomposition Method and Nuclear Rainbow}
\label{sec:NF-standard}

Nuclear rainbow phenomenon is associated with a broad Airy oscillation pattern observed in elastic 
scattering of $\alpha$ particles and light heavy ions at medium and large angles. Such a rainbow pattern
arises when the absorption is weak and the deep real OP attractively deflects the scattering wave at 
subsurface impact parameters, leading to an interference pattern analogous to that of atmospheric rainbow. 
This pattern comprises the first Airy minimum $A_1$ followed by a shoulder-like ``rainbow maximum'', 
and higher-order Airy minima $A_2, A_3...$ \cite{Bra97,Kho07r}. Extensive elastic scattering data 
measured for several light systems such as \OO and \CC in the energy range $E_{\rm lab}\sim 10--70$~MeV/nucleon reveal the evolution of the Airy oscillation pattern with energy, 
and help to constrain the radial shape of the real OP at short distances where the nuclear overlap 
density is reaching up to twice the saturation density \cite{Kho07r}. 

\begin{figure}[tb]
	\centering\vspace*{1cm}
	\includegraphics[width=0.6\columnwidth]{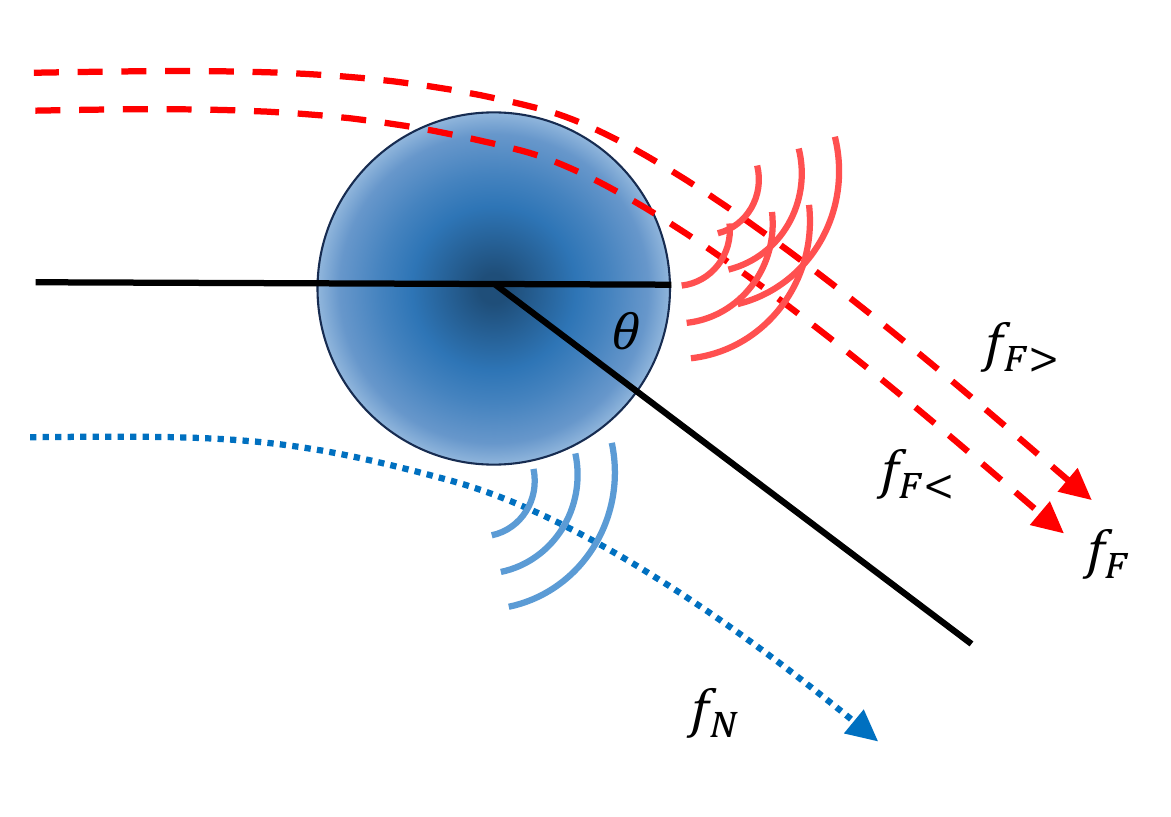}
	\caption{Schematic illustration of semiclassical trajectories of the incident wave scattered by 
	the nuclear OP. Reprinted with permission from Ref.~\cite{Phu24a}.} 	\label{fig:deflection}
\end{figure}

Semiclassically, nuclear rainbow can be characterized by a deflection function $\Theta(L)$ determined 
from the scattering phase shifts, which relates the scattering angle to the orbital momentum $L$ 
\cite{Bra97,Kho07r,Hus84}. The minimum of $\Theta(L)$ at $L_R$ defines the rainbow angle 
$\theta_R=\Theta(L_R)$ that separates the ``bright side'' ($\theta<\theta_R$) from the classically 
forbidden ``dark side'' ($\theta>\theta_R$) of nuclear rainbow. The interference between the outer 
($L>L_R$) and inner ($L<L_R$) branches of trajectories refracted to the same angle generates 
the Airy oscillation pattern of nuclear rainbow (see Fig.~\ref{fig:deflection}) observed in several 
weakly absorbing light HI systems. 
A helpful tool for a quantitative optical model (OM) analysis of elastic \AA scattering is the NF 
decomposition method developed by Fuller~\cite{Ful75,Hus84}. For elastic scattering of two 
nonidentical (spin-zero) nuclei, the total elastic scattering amplitude takes the standard OM form
\begin{equation}
	f(\theta)
	= f_R(\theta) 	+ \frac{1}{2ik} \sum_{L=0}^{\infty} (2L+1)\,
	e^{2i\sigma_L}\,\bigl(S_L-1\bigr)\,P_L(\cos\theta),
	\label{eq:full-amplitude}
\end{equation}
where $k$ and $f_R(\theta)$ are the c.m.\ momentum and Rutherford scattering amplitude, respectively. 
$\sigma_L$ , $S_L$, and $P_L$ are $L$-th partial wave components of the Coulomb phase shift, elastic 
$S$ matrix, and Legendre polynomial, respectively. Fuller split the Legendre function $P_L(\cos\theta)$ 
into two waves scattered at $\theta$ but running in the opposite directions around the scattering center 
as illustrated in Fig.~\ref{fig:deflection}, using the Legendre function of the second kind $Q_L$ 
\begin{equation}
	\tilde Q_L^{(\mp)}(\cos\theta)	= \frac{1}{2}\left[P_L(\cos\theta) \pm \frac{2i}{\pi}
	Q_L(\cos\theta)\right]. \label{eq:Qtilde-def}
\end{equation}
Because the Coulomb scattering amplitude is singular at $\theta=0$, where the partial-wave series 
diverges, Fuller projected the argument of the Legendre function into the complex plane to get around 
the singularity, and obtained the NF components of the Rutherford scattering amplitude in a closed 
analytical form \cite{Ful75}. The total elastic scattering amplitude can then be decomposed into 
the NF components as
\begin{align}
	f(\theta) 	&= f_N(\theta) + f_F(\theta), \label{eq:f-NF-split}\\[1ex]
	f_N(\theta)	&= f_R^{(N)}(\theta)	+ \frac{1}{2ik} \sum_L (2L+1)\,
	e^{2i\sigma_L}\,\bigl(S_L-1\bigr)\,	\tilde Q_L^{(-)}(\cos\theta), \label{eq:f-N}\\[1ex] 
	f_F(\theta) 	&= f_R^{(F)}(\theta)	+ \frac{1}{2ik} \sum_L (2L+1)\,
	e^{2i\sigma_L}\,\bigl(S_L-1\bigr)\,\tilde Q_L^{(+)}(\cos\theta). \label{eq:f-F}
\end{align}
A nice feature of the scattering theory is that the $L$-th partial wave associated with 
$\tilde Q_L^{(-)}(\cos\theta)$ is deflected from the \emph{near side} of the scattering center to angle 
$\theta$, and that associated with $\tilde Q_L^{(+)}(\cos\theta)$ is deflected from the opposite, 
\emph{far side} of the scattering center to the same angle $\theta$ (see Fig.~\ref{fig:deflection}). 
As a result, the nearside amplitude $f_N$ describes mainly the repulsive diffraction-like surface 
scattering, while the farside amplitude $f_F$ accounts for the attractively refracted wave, which 
penetrates more into the nuclear interior. Since the Coulomb scattering is repulsive, mainly surface 
scattering, $f_R^{(F)}(\theta)$ is much weaker than the nuclear farside amplitude that shapes 
the Airy oscillation pattern in the farside scattering cross section 
$\mathrm{d}\sigma_F/\mathrm{d}\Omega = |f_F(\theta)|^2$. The location of the primary Airy 
minimum $A_1$ and the extension of the Airy pattern into the dark side of nuclear rainbow 
are mainly determined by the strength and shape of the real OP at small radii \cite{Kho07r}. 
Reproducing $\mathrm{d}\sigma_F/\mathrm{d}\Omega$ over several orders of magnitude 
provides, therefore, a strong constraint on the real OP. 
\begin{figure}[tb]\vspace*{-1cm}
	\centering
	\includegraphics[width=1.1\columnwidth]{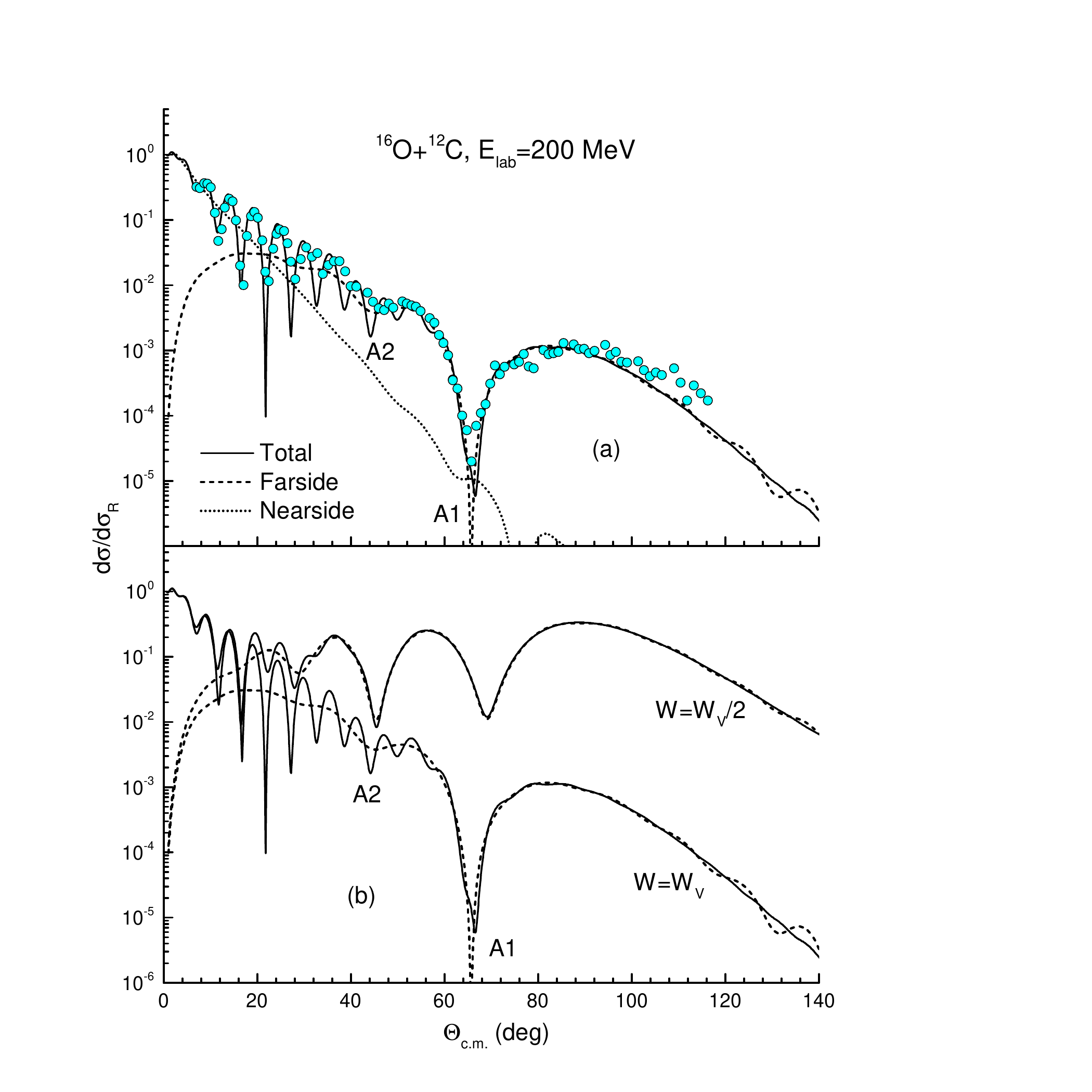}
\caption{(a) OM description of elastic \OC scattering data at $E_{\rm lab}=200$ MeV 
\cite{Oglob00} given by the folding model analysis of Ref.~\cite{Kho16}. The total elastic 
cross section (solid line), nearside (dotted line) and farside (dashed line) contributions
were obtained using Eqs.~(\ref{eq:f-NF-split}), (\ref{eq:f-N}) and (\ref{eq:f-F}), respectively. 
(b) Total elastic (solid line), and farside (dashed line) cross sections were obtained with 
two absorptive strengths of the imaginary OP. A1 and A2 denote the first- and second-order Airy 
minima, respectively. } 	\label{f2}
\end{figure}

A typical example is shown in Fig.~\ref{f2} where elastic \OC scattering data measured
at $E_{\rm lab}=200$ MeV by Ogloblin {\it et al.} \cite{Oglob00} are compared with the OM 
results given by the folding model analysis of Ref.~\cite{Kho16}. The total elastic cross section
(solid line), the nearside (dotted line) and farside (dashed line) cross sections were obtained using 
Eqs.~(\ref{eq:f-NF-split}), (\ref{eq:f-N}) and (\ref{eq:f-F}), respectively. This is among the most 
prominent cases of nuclear rainbow observed so far, where the first Airy minimum A1 (at 
$\theta_{\rm c.m.}\approx 64^\circ$) followed by a shoulder-like rainbow maximum is entirely 
given by the farside scattering. The second Airy minimum A2 of nuclear rainbow can be identified 
in measured elastic data at $\theta_{\rm c.m.}\approx 43^\circ$. Given the farside 
scattering cross section determined mainly by the real OP, the OM calculation done with the same 
real OP, but with the absorptive strength of the imaginary OP decreased by 50\%, shows very clearly 
the location of A1 and A2. The OM results plotted in Fig.~\ref{f2} illustrate nicely how the Airy 
oscillation pattern of nuclear rainbow can be revealed by the Fuller decomposition of elastic 
scattering amplitude into the NF components.     

It is noteworthy that the barrier-internal (BI) decomposition of elastic scattering wave suggested 
by Brink and Takigawa \cite{Bri77,Row77,Alb82} offers an alternative description of the Airy 
oscillation pattern by coherently summing amplitudes of wave reflected at- and wave penetrating 
through the Coulomb + centrifugal barrier \cite{Mi00}. Both the NF and BI decomposition methods 
identify nuclear rainbow as a refractive scattering phenomenon that disappears when the strong 
absorption damps the subsurface trajectories. 

\section{NF Decomposition Method for Systems with Core-Exchange Symmetry}
\label{sec:core-exchange}

In systems containing two identical cores, the scattering amplitude is constrained by the core-exchange 
symmetry, even when projectile and target are not identical. The core-exchange symmetry is most 
transparent in elastic scattering of two identical (spin-zero) nuclei such as \CC and \OO, where the 
boson symmetry implies the total scattering wave function to be symmetric under the projectile-target 
exchange (see Fig.~\ref{f3}), which leads to the well-known Mott interference pattern in the angular
distribution. In a core-identical system like \OC or \CCC, a similar interference arises from the elastic 
transfer of valence cluster or nucleon between two identical cores, which can be effectively encoded 
in a parity-dependent OP. This section presents the NF decomposition method generalized to properly
account for the core-exchange symmetry, and what insights emerge from the nuclear rainbow and 
large-angle oscillation of elastic scattering cross section.

\subsection{Identical systems}
\begin{figure}[tb]
	\centering
	\includegraphics[width=0.9\columnwidth]{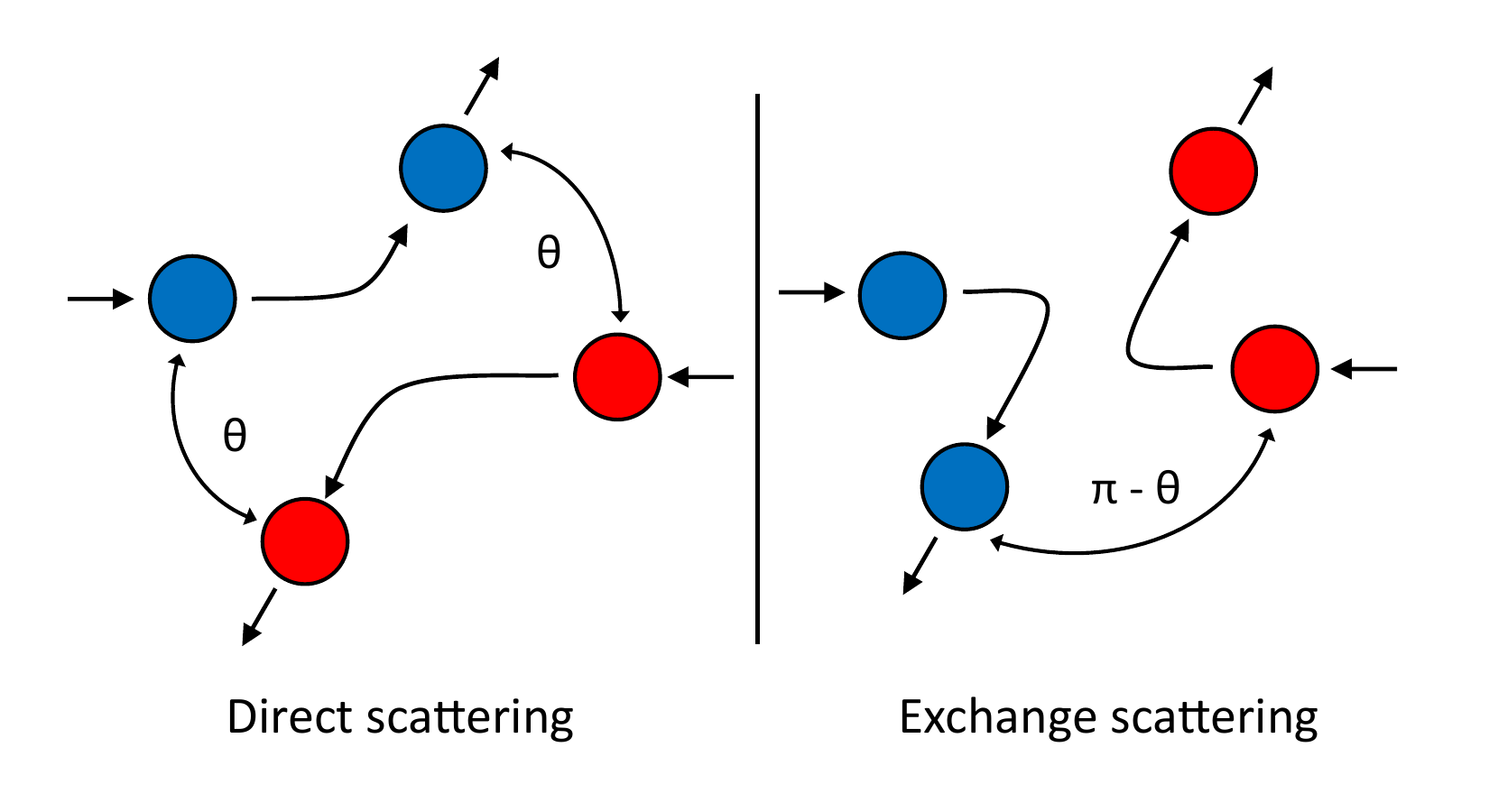}
	\caption{Schematic illustration of the direct scattering (at angle $\theta$) and exchange scattering 
	(at angle $\pi-\theta$) for an identical dinuclear system. Reprinted with permission from Ref.~\cite{Phu24a}.} \label{f3}
\end{figure}

The elastic scattering (ES) wave function of two identical spin-zero nuclei must be symmetric 
with respect to the projectile-target exchange, so that the total ES amplitude is also symmetric 
and given in terms of the Mott amplitude and the symmetrized nuclear amplitude \cite{Phu24a} as
\begin{equation}
	f_{\rm ES}(\theta)	= f_{\rm Mott}(\theta) + f_{\rm sym}(\theta),\ \mbox{with}\ 
	f_{\rm sym}(\theta)=f(\theta) + f(\pi-\theta).
\end{equation}
Here $\theta\equiv\theta_{\rm c.m.}$ and $f_{\rm Mott}(\theta)=f_R(\theta)+f_R(\pi-\theta)$, 
where $f_R$ is the Rutherford scattering amplitude. The unsymmetrized $f_R(\theta)$ and $f(\theta)$ 
amplitudes account for the direct scattering contribution, while $f_R(\pi-\theta)$ and $f(\pi-\theta)$ 
account for the (recoiled) exchange scattering contribution, as illustrated in Fig.~\ref{f3}.  

Applying Fuller's method \cite{Ful75} to separately decompose $f_R(\theta)$ and $f_R(\pi-\theta)$ 
into the NF components by continuing the partial-wave series into the complex plane, the NF 
components of the Mott amplitude are rigorously obtained \cite{Phu24a} as
\begin{equation}
	f_{\rm Mott}^{(N)}(\theta) = f_R^{(N)}(\theta)+f_R^{(F)}(\pi-\theta)\ \ \mbox{and}\ \ 
	f_{\rm Mott}^{(F)}(\theta) = f_R^{(F)}(\theta)+f_R^{(N)}(\pi-\theta), \label{eqNF_Mott}
\end{equation}
where $f_R^{(N/F)}$ are the NF components of the Rutherford amplitude derived in analytical form 
by Fuller \cite{Ful75}. As can be seen in Eq.~\eqref{eqNF_Mott}, the nearside component of the 
Mott amplitude is a superposition of the nearside and farside components of the Rutherford amplitude 
determined at the angles $\theta$ and $\pi-\theta$, respectively, and vice versa for the farside 
component of the Mott amplitude. $f^{(N)}_{\rm Mott}(\theta)$ and $f^{(F)}_{\rm Mott}(\theta)$ 
become equal at $\theta=\pi/2$.

The symmetrized nuclear scattering amplitude can be decomposed into the nearside and farside 
components by expressing the partial-wave series in terms of Legendre functions of the second kind
\begin{align} 
f_{\rm sym}(\theta)=f^{(N)}_{\rm sym}(\theta)+f^{(F)}_{\rm sym}(\theta),\hspace{4cm} \\
 =\frac{1}{2ik}\sum_L (2L+1)e^{2i\sigma_L}(S_L-1)\left[1+(-1)^L\right]\left[\tilde Q_L^{(-)}(\cos\theta)
 + \tilde Q_L^{(+)}(\cos\theta)\right], \label{eqNF_nuc}
\end{align}
where the summation is done over \emph{even} partial waves $L$ only. Given the relation\linebreak 
$(-1)^L\tilde Q_L^{(\pm)}(\cos\theta)=\tilde Q_L^{(\mp)}(\cos(\pi-\theta))$, the nearside and 
farside components of the symmetrized nuclear amplitude (\ref{eqNF_nuc}) are readily obtained as
\begin{equation}
	f_{\rm sym}^{(N)}(\theta)=f^{(N)}(\theta)+f^{(F)}(\pi-\theta)\ \ \mbox{and}\ \
	f_{\rm sym}^{(F)}(\theta)=f^{(F)}(\theta)+f^{(N)}(\pi-\theta).
\end{equation}
As a result, the nearside component $f_{\rm sym}^{(N)}$ of the symmetrized nuclear scattering 
amplitude is also a superposition of the nearside and farside components of the unsymmetrized 
nuclear amplitude at angles $\theta$ and $\pi-\theta$, respectively, and vice versa for 
$f_{\rm sym}^{(F)}$, in the same way as for the Mott scattering amplitude.

\begin{figure}[bt]\vspace*{0cm}\hspace*{0cm}\centering
\includegraphics[width=0.8\textwidth]{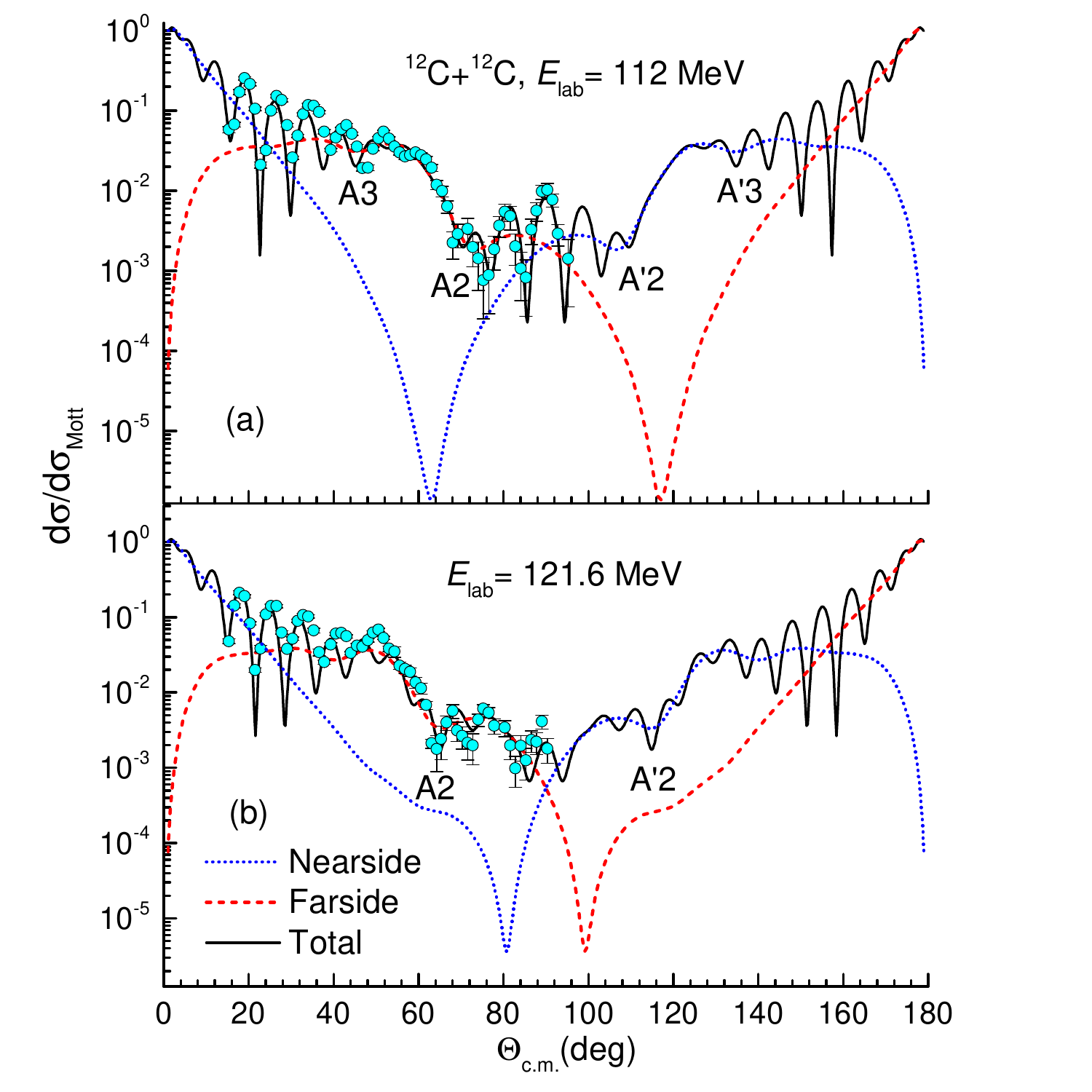}\vspace*{0cm}
 \caption{(a) Elastic \CC scattering data measured at $E_{\rm lab}=112$ MeV \cite{Sto79} 
in comparison with results of the OM calculation, taking exactly into account the projectile-target 
symmetrization (solid line). The nearside (dotted line) and farside (dashed line) cross sections were 
given by the NF decomposition (\ref{eqDX2}) of the ES amplitude. A$n$ is the $n$th-order Airy 
minimum at angle $\theta<90^\circ$, and A'$n$ is its symmetric partner at angle $180^\circ-\theta$. 
(b) The same as (a) but for the elastic \CC scattering data measured at 
 $E_{\rm lab}=121.6$ MeV \cite{Sto79}. Reprinted with permission from Ref.~\cite{Phu24a}.} \label{f4}
\end{figure}

It is helpful to express the total ES amplitude in terms of the direct (D) and exchange (EX) 
scattering amplitudes $f_{\rm ES}(\theta)=f_{\rm D}(\theta)+f_{\rm EX}(\pi-\theta)$ 
\begin{equation}
\mbox{where}\ f_{\rm D}(\theta)=f_R(\theta)+f(\theta),\  
\mbox{and}\ f_{\rm EX}(\pi-\theta)=f_R(\pi-\theta)+f(\pi-\theta). 	\label{eqDX1}
\end{equation} 
The consequence of relation (\ref{eqDX1}) is a symmetric interchange of the direct and exchange 
scattering amplitudes $f_{D}\rightleftarrows f_{EX}$ as the scattering angle $\theta$ 
passes through $90^\circ$. Combining the NF components of the Mott and nuclear scattering 
amplitudes, the total elastic NF amplitudes are obtained as 
\begin{equation}
	f_{\rm ES}^{(N)}(\theta)=f_{\rm D}^{(N)}(\theta)+f_{\rm EX}^{(F)}(\pi-\theta)\ \mbox{and}\ 
	f_{\rm ES}^{(F)}(\theta)=f_{\rm D}^{(F)}(\theta)+f_{\rm EX}^{(N)}(\pi-\theta). \label{eqDX2}
\end{equation}
In a similar manner, the relations (\ref{eqDX2}) imply a symmetric interchange of the nearside 
and farside components of the total ES amplitude $f^{(N)}_{\rm ES}\rightleftarrows f^{(F)}_{\rm ES}$ 
as the scattering angle $\theta$ passes through $90^\circ$. For strongly refractive, symmetric 
systems like \CC and \OO, the resulting NF scattering cross sections exhibit a characteristic 
``butterfly-wing" pattern \cite{Phu24a} where each Airy minimum of the direct farside cross section 
at angle $\theta<90^\circ$ has its symmetric partner in the exchange farside cross section at angle
$180^\circ-\theta$ as shown, e.g., for \CC system in Fig.~\ref{f4}.

One can see in Fig.~\ref{f4} that the elastic cross section is symmetric about $\theta=90^\circ$, and
the nearside and farside cross sections at angles $\theta<90^\circ$ are symmetrically interchanged 
to the farside and nearside cross sections at angles $\theta>90^\circ$, respectively. Thus, 
the symmetrization of the ES amplitude results in a symmetric Airy oscillation pattern, and each 
Airy minimum located at angle $\theta<\pi/2$ has its symmetric partner located at angle $\pi-\theta$. 
Given a weak (direct) nearside cross section at large angles \cite{Phu24a}, the quickly oscillating ES 
cross section around $90^\circ$ is in fact an interference of two farside amplitudes (the direct and 
exchange ones) generated by the same OP. As a result, the ES cross section around $\theta=90^\circ$ 
becomes quite sensitive to the shape of the real OP at sub-surface distances (see more details
in Ref.~\cite{Phu24a}). 

In summary, the generalized NF decomposition method is very helpful for the study of the subtle rainbow 
pattern resulting from the boson symmetry of an identical (spin-zero) dinuclear system. 

\subsection{Core-identical systems}
\begin{figure}[tb]
	\centering
	\includegraphics[width=0.9\columnwidth]{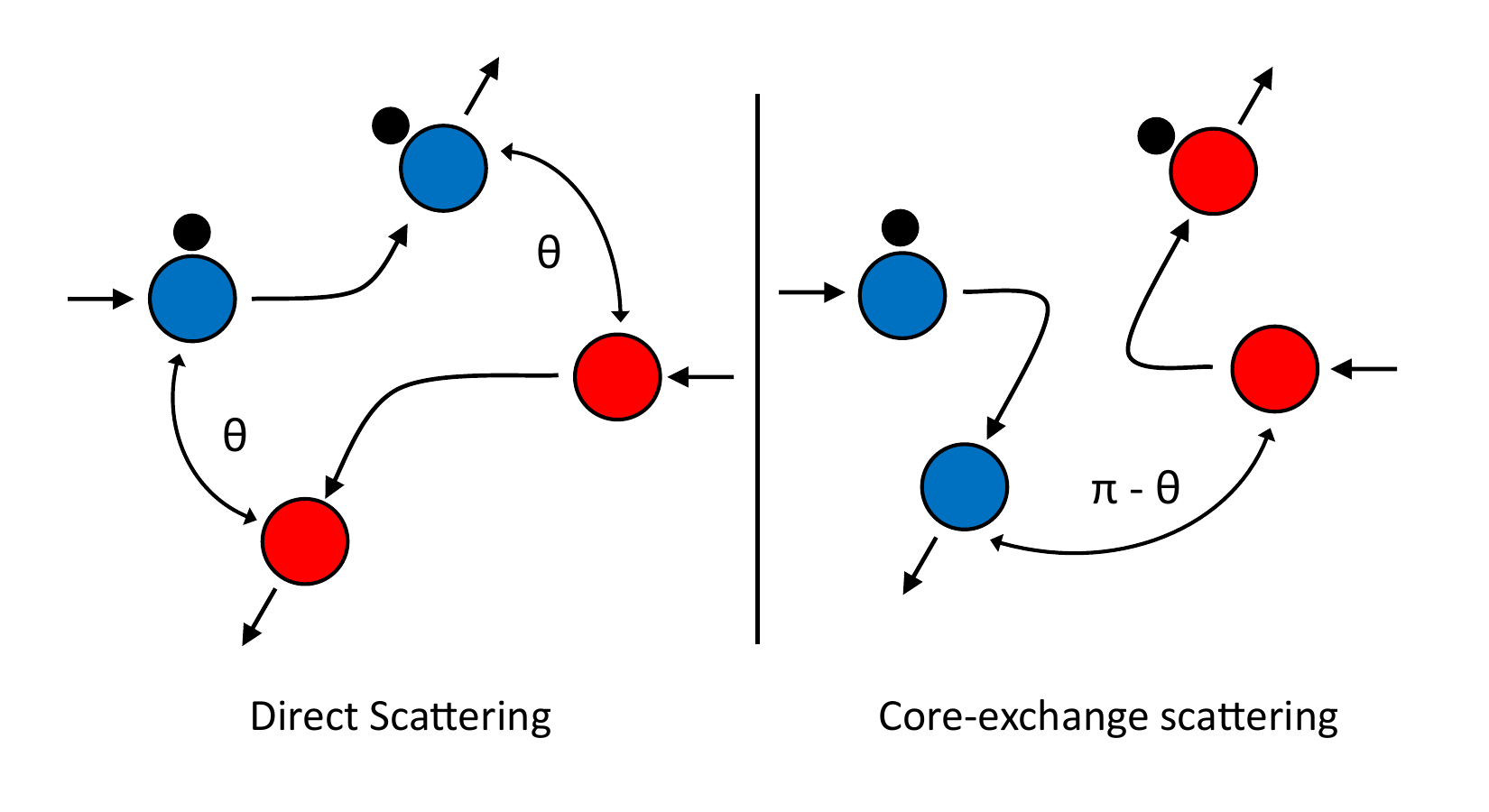}
	\caption{Elastic transfer viewed as the exchange of two identical cores. The red and blue spheres 
	are two cores in the initial states of projectile and target, and the black circle is the valence nucleon or cluster being transferred. Reprinted with permission from Ref.~\cite{Phu19}.} \label{f5}
\end{figure}

In a core-identical system $(A{+}x)+A$, such as \OC or \CCC, both the elastic scattering channel 
$A(A{+}x,A{+}x)A$ and the elastic transfer channel $A(A{+}x,A)(A{+}x)$ have the same final 
configuration, which cannot be distinguished experimentally (see Fig.~\ref{f5}). Therefore, the total 
elastic amplitude is given as \cite{Phu18,Phu19}
\begin{equation}
	f_{\rm total}(\theta) = f_{\rm ES}(\theta)+f_{\rm ET}(\pi-\theta),
\end{equation}
where $f_{\rm ES}$ and $f_{\rm ET}$ are the elastic scattering and elastic transfer amplitudes, 
respectively. In general, the ES and ET channels must be treated separately in the coupled reaction 
channels (CRC) calculation \cite{Tho88,vOe75}. 

The elastic transfer is physically equivalent to the core-exchange process as illustrated in Fig.~\ref{f5}, so that $f_{\rm ET}(\pi-\theta)$ is analogous to the exchange scattering amplitude $f_{\rm EX}(\pi-\theta)$ 
considered above for an identical system. For two (spin-zero) identical cores, the total elastic amplitude 
can be expanded in the partial wave series \cite{Phu19,Phu24a} as 
\begin{equation}
 f_{\rm total}(\theta)=f_R(\theta)+\frac{1}{2ik}\sum_L(2L+1)e^{2i\sigma_L}
 \left[S_{\rm ES}^{(L)}+(-1)^L S_{\rm ET}^{(L)}-1\right]P_L(\cos\theta), \label{eqEST}
\end{equation}
where $f_{\rm ES}$ and $f_{\rm ET}$ are given by the CRC solutions obtained for the ES and 
ET channel wave functions, respectively. The summation of the partial wave series (\ref{eqEST}) is done over 
both \emph{odd} and \emph{even} partial waves $L$. Our recent CRC calculations of elastic $\alpha$ 
transfer in the \OC system \cite{Phu18,Phu19,Phu21a,Phu24a} have shown that $f_{\rm ET}$ strongly 
enhances the elastic cross section at backward angles, with a rapid oscillation caused by the interference 
of $S_{\rm ES}$ and $S_{\rm ET}$ that obscures the nuclear rainbow pattern. Although the final state 
of the ES and ET channels is the same, there is no way to rigorously link $S_{\rm ES}$ and 
$S_{\rm ET}$ matrices to the same OP. While $S_{\rm ES}$ is associated with elastic (Coulomb+nuclear) 
scattering, $S_{\rm ET}$ represents the elastic $\alpha$ transfer between two $^{12}$C cores, which 
is associated with the dissociation $^{16}$O$\to\alpha+^{12}$C. 

\begin{figure}\vspace*{-0.5cm}\hspace*{0cm}\centering
\includegraphics[width=0.8\textwidth]{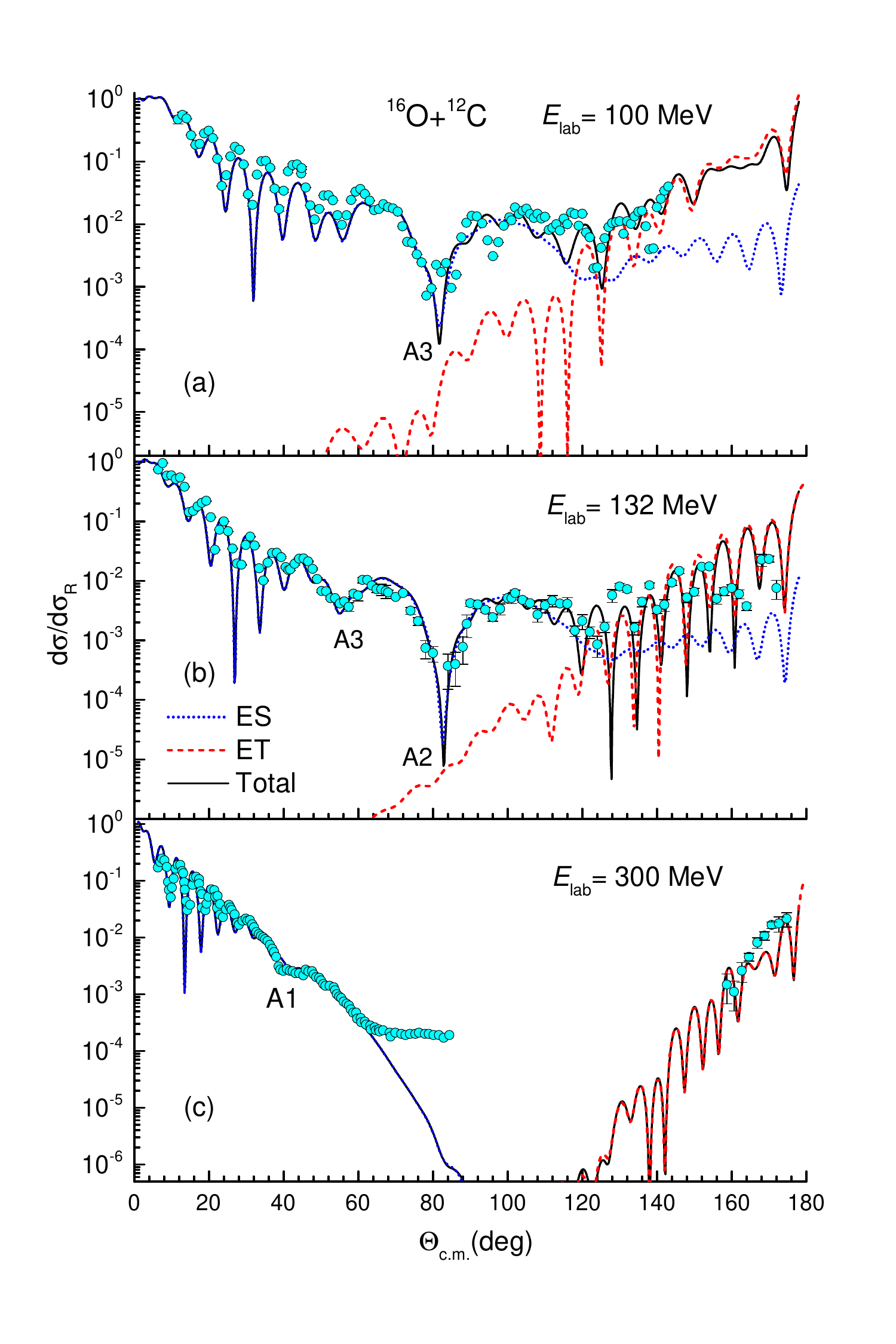}\vspace*{-1cm}
 \caption{Two-channel CRC description of elastic \OC data measured at $E_{\rm lab}=100$ MeV (a), 
132 MeV (b), and 300 MeV (c) \cite{Oglob00,Nico00,Oglo98,Bra01}, where the ES, ET, and total (ES+ET) 
elastic cross sections are shown as dotted, dashed, and solid lines, respectively. A$n$ is the $n$th-order Airy 
minimum of the ES cross section. Reprinted with permission from Ref.~\cite{Phu24a}.} \label{f6}
\end{figure}
The elastic \OC cross sections obtained from the total elastic amplitude (\ref{eqEST}) given by the two-channel 
CRC calculation are compared with elastic \OC data measured at $E_{\rm lab}=100$ MeV \cite{Nico00}, 
132 MeV \cite{Oglob00,Oglo98}, and 300 MeV \cite{Bra01} in Fig.~\ref{f6}, and the enhanced 
oscillation of elastic cross section at backward angles is shown to be due mainly to the ET or core-exchange 
process. Although the rainbow pattern at backward angles is distorted by the ET process, the location and 
shape of the Airy minima are still well observed at medium angles.  

A well-known approximation to account for the ET or core-exchange process in a single-channel OM 
calculation is the introduction of a parity-dependent term to the total OP \cite{vOe75,Fra80}. In fact, it is the factor $(-1)^L$ of the $L$-th element of elastic transfer matrix $S_{\rm ET}$ \eqref{eqEST} that
gives rise to a parity-dependent (Majorana) term often introduced to the OP of a core-identical system. 
For example, the inversion of $S_{\rm ET}$ matrix given by the CRC calculation of elastic \OC scattering 
at low energies \cite{Phu19} yields a local OP with a sizable complex Majorana term $V_M$, so that 
the single-channel OM calculation reproduces the full CRC elastic cross section, including the oscillating 
enhancement at backward angles. 

We note that the ES and ET contributions to the total elastic cross section should not be symmetrically 
equal as the direct and exchange scattering amplitudes discussed above for two identical nuclei. 
To explore the rainbow pattern in more detail, the NF decomposition has been done separately for the 
ES and ET amplitudes \cite{Phu24a}, and the NF components of the total elastic amplitude can be 
expressed as
\begin{equation}
	f_{\rm tot}^{(N)}(\theta)=f_{\rm ES}^{(N)}(\theta)+f_{\rm ET}^{(F)}(\pi-\theta)\ 
	\mbox{\rm and}\ 
	f_{\rm tot}^{(F)}(\theta)=f_{\rm ES}^{(F)}(\theta)+f_{\rm ET}^{(N)}(\pi-\theta), \label{eqNFT}
\end{equation}
which mirrors the identical-particle structure \eqref{eqDX2} but without exact NF symmetry, because 
the ES and ET strengths differ and there is no Rutherford amplitude in the ET channel. 
\begin{figure}\vspace*{0cm}\hspace*{0cm}\centering
\includegraphics[width=0.8\textwidth]{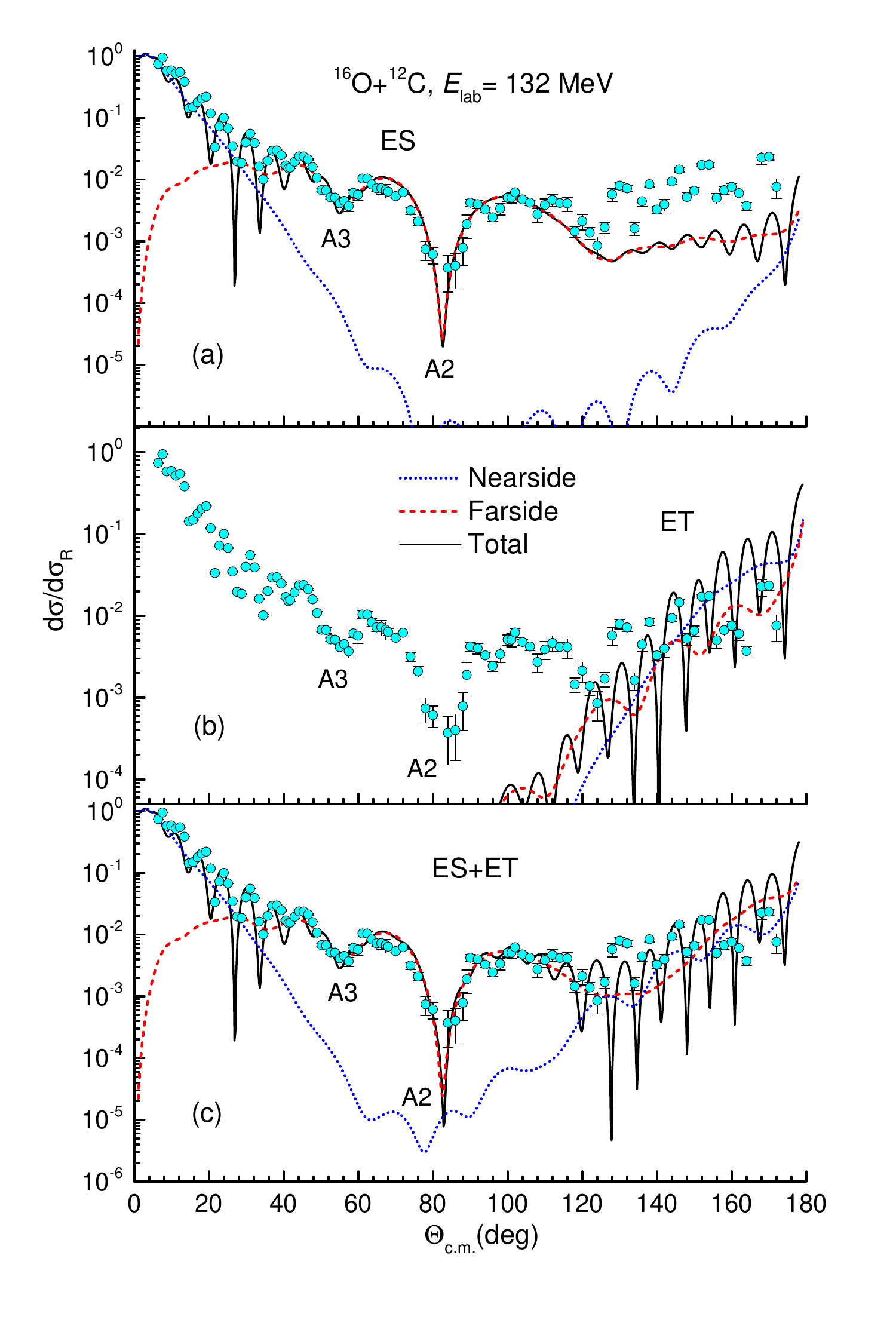}\vspace*{-1cm}
 \caption{NF decomposition of the elastic \OC amplitude at $E_{\rm lab}=132$ MeV given
by the two-channel CRC calculation: (a) - for the ES only, (b) - for the ET only, and (c) - for the total 
(ES+ET) elastic amplitude. Reprinted with permission from Ref.~\cite{Phu24a}.} \label{f7}
\end{figure}
The NF decomposition was done separately for the ES, ET, and total (ES+ET) elastic amplitudes 
\eqref{eqNFT} determined by the two-channel CRC calculation of elastic \OC scattering 
at $E_{\rm lab}=132$ MeV, and the results are shown with measured data \cite{Oglob00,Oglo98} 
in Fig.~\ref{f7}. One can see that the ET cross section (b) is \emph{not} a symmetrical reflection 
of the ES cross section (a). While the ES cross section is dominated by the farside scattering over a wide 
angular range, the ET cross section is a typical \emph{diffractive} interference pattern, which indicates 
a surface character of the ET process. In fact, the ``recoiled" contribution of $f_{\rm ET}(\pi-\theta)$ 
to $f_{\rm total}(\theta)$ at backward angles represents the ET occurring at forward angles as illustrated 
in Fig.~\ref{f5}.

\begin{figure}[tb]\hspace*{-0.5cm}\centering
	\includegraphics[width=1.1\columnwidth]{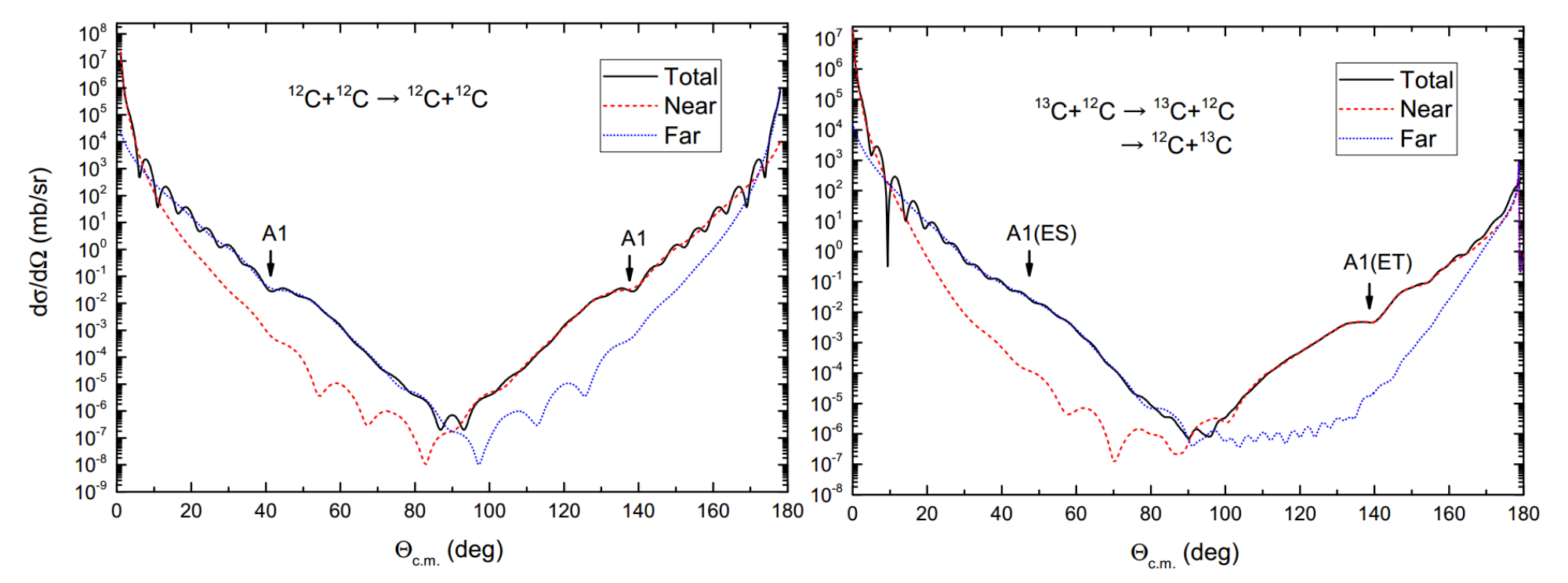}
	\caption{NF decomposition of elastic scattering for \CC (left) and \CCC (right) at $E_{\rm c.m.}=120$ MeV. 
The total cross section, its nearside and farside components are shown as solid, dashed, and dotted lines, 
respectively. For the \CC system, the symmetric butterfly-wing pattern exhibits the first Airy minimum 
$A_1$ and its mirrored image at $\theta$ and $\pi-\theta$, respectively. For the \CCC system, $A_1$ minimum 
of ES cross section and that of elastic neutron transfer appear at slightly different angles, but still reflecting 
the NF interchange pattern resulted from the core-exchange symmetry.} \label{f8}
\end{figure}
Our further study of elastic \CCC scattering \cite{Phu26} indicates that the same core-exchange mechanism 
operates when a valence neutron is added to one of the two $^{12}$C cores. By introducing a parity-dependent 
Majorana term constrained by the CRC analysis to the OP and applying the generalized NF decomposition 
method, one finds that the NF pattern of elastic $^{13}$C+$^{12}$C cross section evolves smoothly from 
that of the symmetric \CC system (Fig.~\ref{f8}). Thus, the core-exchange symmetry, whether exact or approximate, 
provides a unifying framework for the revelation of the NF scattering pattern of nuclear rainbow, and large-angle scattering structures observed for the refractive light HI systems.

\section{Summary and Outlook}
\label{sec:summary}

In this contributing paper we have summarized the recent extension of the NF decomposition method of light HI 
scattering into a unified framework for the analysis of elastic (refractive) light HI scattering, including systems 
with core-exchange symmetry. Based on Fuller's original technique, this formalism isolates the farside scattering 
amplitude responsible for the Airy oscillation pattern of nuclear rainbow. 

For an identical spin-zero system like \CC or \OO, the boson symmetry leads to the NF interchange at 
$\theta_{\rm c.m.}=90^\circ$, with a characteristic butterfly-wing Airy pattern of the farside cross 
section \cite{Phu24a}. In a core-identical system like \OC or \CCC, a coherent superposition of elastic 
scattering and elastic ($\alpha$ or nucleon) transfer amplitudes generates an analogous NF interchange 
at angles around $90^\circ$, encoding both the subsurface refraction and surface-dominated core-exchange 
dynamics~\cite{Phu18,Phu19}. This substantially enhances the sensitivity of elastic data taken at medium 
and large angles to the nucleus-nucleus OP.

Future development should interface the generalized NF analysis with the nonlocal OP obtained from 
the microscopic calculations, enabling refractive observables to validate the few-body description 
of the nuclear interaction. Extension to explicit few-body projectiles and breakup channels would 
also strengthen links between nuclear rainbow scattering and nuclear-data applications. Combined 
with modern microscopic inputs, the generalized NF analysis of nuclear rainbow scattering could offer a path to constrain the real OP, probe cluster correlations, and enable a deeper understanding 
of the few-body dynamics.

\begin{acknowledgements}
The present research has been supported, in part, by the National Foundation for Science and Technology 
Development of Vietnam (NAFOSTED Project No. 103.04-2025.06).
\end{acknowledgements}



\end{document}